\documentclass[twocolumn,showpacs,preprintnumbers,amsmath,amssymb]{revtex4}

\usepackage{amsmath}
\usepackage{graphicx}
\usepackage{dcolumn}
\usepackage{bm}

\begin{document}

\preprint{}

\title{Metamaterial-based polarization control plate for producing incoherent laser irradiation}
\author{Xiaohui Ling}
\author{Hailu Luo}
\author{Chujun Zhao}
\author{Shuangchun Wen}\email{scwen@hnu.edu.cn}
\author{Dianyuan Fan}
\affiliation{Key Laboratory for Micro-/nano Optoelectronic Devices
of Ministry of Education, College of Information Science and
Engineering, Hunan University, Changsha 410082, People's Republic of
China}
\date{\today}

\begin{abstract}
We present a metamaterial-based random polarization control plate to
produce incoherent laser irradiation by exploiting the ability of
metamaterial in local polarization manipulation of beam upon
transmission via tuning its local geometry. As a proof-of-principle,
we exemplify this idea numerically in a simple optical system using
a typical L-shaped plasmonic metamaterial with locally varying
geometry, from which the desired polarization distribution can be
obtained. The calculating results illustrate that this scheme can
effectively suppress the speckle contrast and increase irradiation
uniformity, which has potential to satisfy the increasing
requirements for incoherent laser irradiation.

\end{abstract}


\maketitle

\section{Introduction}\label{SecI}
Coherence is a fundamental property of laser, which facilitate laser
to be widely applied in those fields requiring high coherent light
sources. However, just as a coin has two sides, coherence is harmful
to some laser application fields. High coherence usually results in
unwanted speckle noise in laser display (or laser projection imaging
or laser TV), thereby decreasing the imaging
resolution~\cite{Mandel1995}. In laser fusion and laser heat
processing, coherence makes the intensity distribution on the focal
plane not uniform enough~\cite{Deng1986}. In some spectroscopy
experiments, high coherence laser is also undesirable. Thus, in
these applications, it is urgently need to eliminating the laser
coherence.

In essence, controlling the polarization or phase of laser beam
could achieve incoherent laser irradiation. The key points of beam
docoherence through polarization control can be summarized as
follows. It first breaks a laser beam into two or many beamlets with
orthogonal or random polarization distribution and then converge
them; the irradiation on the focal plane is thereby incoherent. Some
concrete schemes have been proposed, such as using a birefringent
wedge to create two orthogonally polarized beams with a selected
angular separation and employing an optic made by wave plates to
scramble the polarization distribution in the near
field~\cite{LLE1990,Tsubakimoto1992,Rothenberg2000,Munro2004}.
Generally, these polarization control devices are made of nematic
liquid crystal, KDP crystals, and crystal half-wave plates, etc.

On the other hand, as artificial periodic subwavelength structures,
metamaterial can exhibit strong anisotropy by suitably designing its
microscopic structure unit~\cite{Smith2004,Hao2007}. Generally,
naturally existing anisotropic mediums exhibit very small
birefringent index in optical domain, which need relatively thick
slabs to achieve a polarization control device (e.g., hundreds or
thousands of wavelengths for a half-wave plate). However, it is
possible for metamaterial to achieve a polarizer with the same
polarization conversion efficient as convention anisotropic mediums,
with its thickness less than a wavelength and transverse dimension
in the order of wavelength
\cite{Rogacheva2006,Liu2007,Chin2008,Gansel2009,Wu2011}. So it holds
great potential in the future nanophotonics applications, and is
particularly amenable to miniaturization. From this point of view,
it would be interesting to apply metamaterial in polarization
control devices.

In this work, a metamaterial-based random polarization control plate
(MRPCP) is proposed for eliminating the laser coherence and
producing incoherent laser irradiation. Locally tailoring the
geometric structure parameters of an anisotropic metamaterial, we
could obtain a locally varying polarization response of laser beam
passing through it, thereby scrambling the polarization state of
beam in the near field and finally eliminating the laser coherence
on the focal plane. Without loss of generality, we will exemplify
our scheme numerically using a variation of the typical L-shaped
metamaterial, although the approach can be applied to an arbitrary
metamaterial geometry offering enough number of free design
parameters. The calculating results show that the MRPCP can
effectively suppress the speckle contrast and increase irradiation
uniformity.

\section{Theory and scheme}\label{SecII}
The mechanism of MRPCP can be considered by employing the multi-beam
interference formula
\begin{eqnarray}
\textit{I}(r,t)&=&\sum_\textit{i}|E_i|^2\nonumber\\
&&+\sum_{i\neq j}2\left[\vec{a_i}\cdot \vec{a_j}\right]
E_iE_j\cos[\Phi_i(r,t)-\Phi_j(r,t)],\label{multi-beam}
\end{eqnarray}
where $\vec{a_{i,j}}$,  $E_{i,j}$, and $\Phi_{i,j}$ represent the
polarization direction vector, field amplitude, and phase of the
($i,j$)th field, respectively. $I(r,t)$ is the total intensity with
location ($r$) and time ($t$) dependence. It is obvious that
incoherent addition of the beamlets is achieved when their
polarization vectors are random, since the vector summation of the
interference term (the second term) in Eq.~(\ref{multi-beam}) is
zero. This is the basis of the polarization docoherence.

\begin{figure}
\includegraphics[height=5cm]{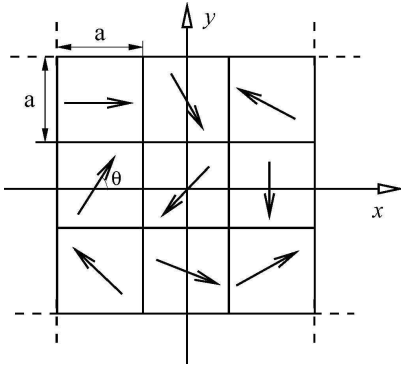}
\caption{\label{Fig1} Schematic picture of the MRPCP. Each square
represents a elements. The arrows indicate $e$-rays of the
anisotropic metamaterial elements.}
\end{figure}
The intensity distribution of MRPCP combined with a random phase
plate (RPP) is referred to as laser
speckle~~\cite{Kato1984,Dixit1993}. The statistical property of the
speckle intensity can be considered as the addition of two
orthogonally polarized speckle patterns (such as in $x$- and
$y$-directions)~\cite{Dainty1984,Goodman2006}. Due to the
non-equivalence of the two speckle intensities ($I_{x}\neq I_{y}$),
the probability density of the intensity can be calculated
by~\cite{Dainty1984,Goodman2006}
\begin{eqnarray}
P_1(I)&=&\frac{1}{\bar{I}_{x}-\bar{I}_{y}}\nonumber\\
&&\times\left[\exp\left(-\frac{I}{\bar{I}_{x}}\right)-\exp\left(-\frac{I}{\bar{I}_{y}}\right)\right],\label{pro}
\end{eqnarray}
with $\bar{I}_{x,y}$ indicating the average intensities of the two
speckle patterns. Because of the incoherency of the two orthogonal
components, we have $\bar{I}_{total}=\bar{I}_{x}+\bar{I}_{y}$, and
then Eq.~(\ref{pro}) can be written as
\begin{eqnarray}
P_1(I)&=&\frac{1}{\bar{I}_{total}(A_1-A_2)}\nonumber\\
&&\times\left[\exp\left(-\frac{I}{\bar{I}_{total}A_1}\right)-\exp\left(-\frac{I}{\bar{I}_{total}A_2}\right)\right],\label{Nonequal}
\end{eqnarray}
where $A_1$ and $A_2$ represent $\bar{I}_{x}/\bar{I}_{total}$ and
$\bar{I}_{y}/\bar{I}_{total}$, respectively. The speckle contrast is
also reduced and determined by the ratio of $A_1$ and $A_2$. The
formula for calculating the contrast is~\cite{Goodman2006}
\begin{eqnarray}
C=\sqrt{A_1^2+A_2^2}/(A_1+A_2).\label{Contrast}
\end{eqnarray}
When $A_1=A_2$, $C$ take the minimum value 0.707. It means that the
speckle contrast can be reduced 29.3$\%$ at most by using MRPCP.

For the convenience of calculation, we now design a MRPCP using
L-shaped plasmonic metamaterial, which requires it producing a
random polarization distribution. A simple scheme is to make the
optic-axis of the metamaterial taking random orientations for
different local regions, namely, locally varying the structure
geometry could create an MRPCP. The schematic picture of the MRPCP
is shown in Fig.~\ref{Fig1} where the arrows indicate the $e$-rays
of the anisotropic metamaterial elements. It is assumed that the
number of the elements is $m\times n$ ($m$ and $n$ are both positive
integers). An element is comprised of several L-shaped unit cells
with the same orientation ($e$-rays) and subwavelength dimension,
and the number of the unit cell for each element is not restricted.

\begin{figure}[b]
\includegraphics[width=8.5cm]{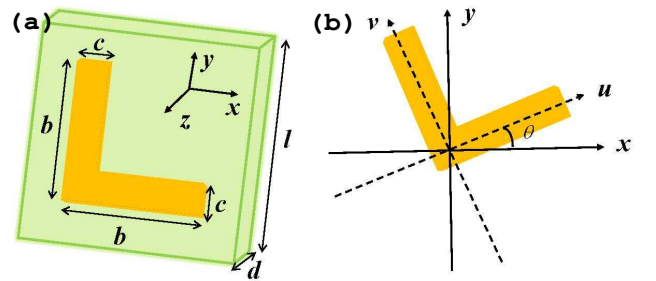}
\caption{\label{Fig2} (a) Schematic of the unit cell with the
geometrical parameters: $b=250$ nm, $c=60$ nm, and the thickness of
the substrate $d=100$ nm. The lattice constant along both $x$- and
$y$-directions are $l=500$ nm. The thickness of the gold ``L'' is
$20$ nm and not shown in this figure. (b) Rotating of ``L'' from the
$x-y$ coordinate system to $u-v$ coordinate system.}
\end{figure}
We first consider the polarization change of an element. The
schematic picture of the metamaterial unit cell is shown in
Fig.~\ref{Fig2}(a). The ``L'' is assumed to be made of gold with 20
nm thickness and built on a dielectric substrate ($n_s=1.45$). They
together form a unit cell of the metamaterial. Other structure
parameters are labeled in the figure. We start by employing the
complex Jones matrix of this metamaterial element~\cite{Menzel2010}.
It can be utilized to describe the transmission of coherent light
through any dispersive optical system. The complex Jones matrix $T$
associating the complex amplitudes of the incident field
($E_{ix,y}$) with the transmitted field ($E_{tx,y}$) is given as
\begin{eqnarray}
\left(\begin{array}{ccc}
 E_{tx}\\
 E_{ty}\end{array}\right) =\left(\begin{array}{ccc}
 T_{xx} & T_{xy}\\
 T_{yx} & T_{yy}\end{array}\right)\left(\begin{array}{ccc}
 E_{ix}\\
E_{iy}\end{array}\right).\label{Jones}
\end{eqnarray}
Here, it is assumed that $\theta=0$ and the incident plane wave
propagates in positive $z$-direction (normal incidence).

Both arms of the ``L'' are mutually perpendicular with identical
structure dimensions. This indicates that $T_{xx}=T_{yy}$ and
$T_{xy}=T_{yx}$. The arms can be seen as oscillators with respective
eigenfrequency, which are associated with the excitation of carriers
representing the free-electron gas of the metal. The carrier
dynamics of one arm is affected by the external electric field as
well as the conductive coupling from another
arm~\cite{Raab2005,Petschulat2010}. Due to the identical structure
dimensions, the eigen resonant responses of the two arms are
identical as well as the conductive coupling responses. So we can
safely obtain that $T_{xx}=T_{yy}$ and $T_{xy}=T_{yx}$.

\begin{figure}
\includegraphics[height=4.0cm]{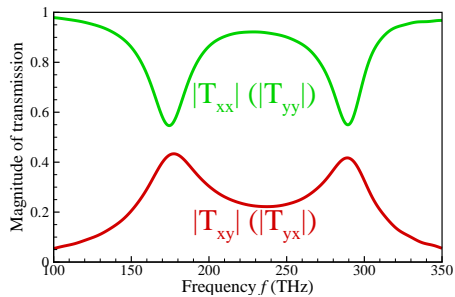}
\caption{\label{Fig3} Magnitude of the transmission coefficient
versus frequency $f$.}
\end{figure}

In order to solve $T_{xx}$ and $T_{xy}$, we use the
finite-difference time-domain method~\cite{Tavlove1995} to simulate
the metamaterial unit cell shown in Fig.~\ref{Fig2}(a). In the
simulations, the metal permittivity is described by Drude model:
$\epsilon(\omega)=1-\omega_p^2/(\omega^2+\textit{i}\gamma\omega)$,
with the bulk plasma frequency $\omega_\textit{p}$ and the
relaxation rate $\gamma$. For gold, from
literature~\cite{Ordal1983}, $\omega_\textit{p}=1.37\times10^4$ THz
and $\gamma=40.84$ THz. The magnitudes of $T_{xx}$ ($T_{yy}$) and
$T_{xy}$ ($T_{yx}$) are plotted in Fig.~\ref{Fig3} when the incident
electric field only polarized in $x$- ($y$-)direction. Each curve
has two valleys (or peaks) which means that each oscillator has two
distinct eigenmodes near the frequencies 173 THz and 290 THz,
respectively~\cite{Petschulat2010}. It also indicates that, if the
incident electric field is polarized in either direction of the two
arms, cross-polarized radiation will generate and of course result
in polarization change in the transmitted field. In fact, the
structure dimensions including the length and thickness of the arms
can affect the resonant responses of the metamaterial, e.g., longer
length or thinner thickness of the arms would decrease the resonant
frequencies.

Now, the Jones vector of the transmitted field can be calculated
according to Eq.~(\ref{Jones}). Without loss of generality, when
$\theta$ takes an arbitrary value [see Fig.~\ref{Fig2}(b)], the
Jones vector can be obtained by transformed the $x-y$ coordinate
system to the $u-v$ coordinate system:
\begin{eqnarray}
\left(\begin{array}{ccc}
 E_{tx}\\
 E_{ty}\end{array}\right)
=R(\theta) \left(\begin{array}{ccc}
 T_{uu} & T_{uv}\\
 T_{vu} & T_{vv}\end{array} \right)R(-\theta)
\left(\begin{array}{ccc}
 E_{ix}\\
E_{iy}\end{array}\right),\label{transform}
\end{eqnarray}
where
\begin{eqnarray}
R(\theta)=\left(\begin{array}{ccc}
 \cos\theta & -\sin\theta\\
 \sin\theta & \cos\theta \end{array} \right).
\end{eqnarray}
According to Eq.~(\ref{transform}), at any frequency, we can obtain
the $\theta$ dependence of the polarization states of the
transmitted field. It's worth noting that the $T$ matrix can be
diagonalized by a rotation of $\pi/4$ or -$\pi/4$ since
$T_{uu}=T_{vv}$ and $T_{uv}=T_{vu}$, which indicates that the angle
bisector direction of ``L'' and its orthogonal direction represent
the $e$- and $o$-rays of the anisotropic metamaterial, respectively.

If both arms of the unit cell are non-identical, $T_{uu}$ is not
equal to $T_{vv}$, but $T_{uv}=T_{vu}$ can still be obtained. The
$T$ matrix could no longer be diagonalized whatever value $\theta$
is. Therefore, actually, we have another degree of freedom to design
the MRPCP, i.e., changing the geometric dimensions of the L's arms.
Arms with different length and thickness will induce different
optical resonance responses (i.e., different $T$ matrix elements).
The $T$ matrix can be simulated for a number of slightly varying
dimensions of ``L'', resulting in a look-up table linking geometric
parameters with $T$ matrix elements. This table represents the set
of possible building blocks of the MRPCP.

\section{Application of the MRPCP for suppressing the speckle contrast and increasing irradiation uniformity}\label{SecIII}
\begin{figure}[b]
\includegraphics[height=4cm]{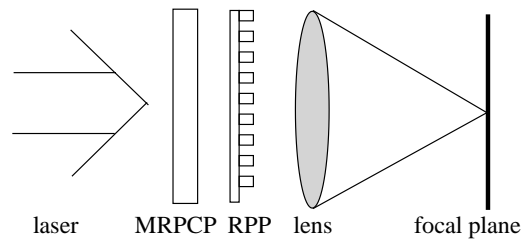}
\caption{\label{Fig4} Schematic picture of the simplified optical
system.}
\end{figure}

As a proof-of-principle for the effectiveness of MRPCP in
suppressing the speckle contrast and increasing irradiation
uniformity, a simple optical system is considered here and
schematically illustrated in Fig.~\ref{Fig4}. A plane wave
$E_{i}(x,y)$ polarized in the $x$-direction ($E_{iy}=0$) with unit
amplitude is assumed for incident laser. It illuminates the MRPCP
and is converged by a lens after passing through a RPP.

According to Eq.~(\ref{transform}), we can obtain the transmitted
field of $x$ and $y$ components from the ($m,n$)th metamaterial
element of the MRPCP:
\begin{eqnarray}
\left(\begin{array}{ccc}
 E_{tx}\\
 E_{ty}\end{array}\right) =\left(\begin{array}{ccc}
 E_{ix}(T_{uu}-T_{uv}\sin2\theta) \\
 E_{ix}T_{uv} \cos 2\theta \end{array} \right).\label{tran}
\end{eqnarray}
From Eq.~(\ref{tran}), it is easily to find that light arising from
different metamaterial elements with random $\theta$ values will
undergo different polarization change. When a laser beam passing
through the random MRPCP, its transverse distribution of
polarization state will be scrambled in the near field and thus make
its spatial coherence become worse. Then, after passing through a
RPP adjacent to the MRPCP, we converge the beamlets by a lens with
focal length $f_0$. As $T_{uu}$ and $T_{uv}$ have been calculated
above by the finite-difference time-domain method, the field
distribution on the focal plane can be derived from the
Fresnel-diffraction integral formula:~\cite{Goodman2005}
\begin{eqnarray}
E_{jmn}(x',y')&=&\frac{\exp(ikf_0)\exp\left[\displaystyle\frac{i
k}{2f_0}(x'^2+y'^2)\right]}{i\lambda f_0}\nonumber\\
&&\times\int\!\!\!\int\limits_{\!\!\!\!\!\!S_{\alpha\beta}}
E_{jmn}(x,y)\exp(i \phi_{RPP})\nonumber\\
&&\times \exp\left[-\frac{i 2\pi}{\lambda f_0}(x x'+y y')\right] dx
dy,
\end{eqnarray}
where $x'$ and $y'$ are axes parallel to the $x$ and $y$ axes,
respectively, $S_{\alpha\beta}$ is the space range of the RPP
illuminated by the $(m,n)$th element of the MRPCP, $\phi_{RPP}$ is
the random phase of the RPP element which is equal to either 0 or
$\pi$, and $j\in(x,y)$. The $j$ component of the total electric
field is given by
\begin{eqnarray}
E_{j}(x',y')=\displaystyle\sum_{m}\displaystyle\sum_{n}E_{jmn}(x',y'),
\end{eqnarray}
and then the intensity of the total electric field can be written as
\begin{eqnarray}
&I_{total}(x',y')&=I_{x}(x',y')+I_{y}(x',y')\nonumber\\
&&=\displaystyle\sum |E_{x}(x',y')|^2+\displaystyle\sum
|E_{y}(x',y')|^2. \label{total}
\end{eqnarray}

\begin{figure}
\includegraphics[width=8.5cm]{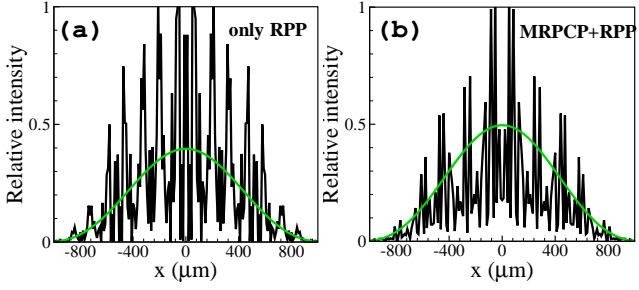}
\caption{\label{Fig5}(Color online) Normalized one-dimensional
intensity distribution (cross section at $y=0$) of the speckle
pattern for (a) only using RPP and (b) MRPCP+RPP. The green solid
lines represent the intensity envelope.}
\end{figure}

Now, we calculate the intensity distribution $I_{total}(x',y')$ on
the focal plane. The parameters used for the calculations are $a=2$
mm, $f_0=1$ m. The operation wavelength is set as 1.5 $\mu$m (i.e.,
$f=200$ THz). The total number of the metamaterial elements is
$16\times16$ whose values of $\theta$ are random numbers between 0
and $2\pi$ produced by the computer. The number of the metamaterial
elements should not be too few so as to fully randomize the
polarization state. The RPP has 64¡Á64 square phase elements (side
length 0.5 mm), and is placed after the MRPCP. The size of the
metamaterial elements are larger than the phase elements in order
that one metamaterial element can illuminate at least one phase
element. The shape and size of the speckle pattern are determined by
those of the individual RPP elements.

The normalized one-dimensional intensity distributions of the
speckle patterns on the focal plane for only using RPP and MRPCP+RPP
are shown in Fig.~\ref{Fig5}. The dense spikes come from the
interference among the beamlets, and the intensity envelope (the
so-called Airy spot indicating by green lines in Fig.~\ref{Fig5}) is
determined by the diffraction of each element of RPP. One can notice
that there is almost no zero-intensity in the central main region
for MRPCP+RPP, while numerous zero-intensities for only using RPP.
Also shown in Fig.~\ref{Fig5}, the introduction of MRPCP
substantially reduces the contrast between the ``spikes'' and
envelope. For further understanding these results, we will
statistically analyze the intensity distributions obtained using
them. The statistical results are calculated by considering
thousands of points of the speckle pattern on the focal plane.

\begin{figure}[b]
\includegraphics[height=4.5cm]{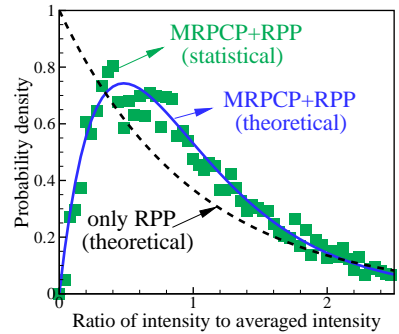}
\caption{\label{Fig6}(Color online) Normalized probability density
($\bar{I}_{total} \times P(I/\bar{I}_{total})$) of intensity
distribution for only using RPP and MRPCP+RPP (theoretical and
statistical).}
\end{figure}
The probability density of intensity for one speckle pattern behaves
as a negative exponential distribution (dashed line in
Fig.~\ref{Fig6}), such as for the case of only using
RPP~\cite{Dixit1993}. For MRPCP+RPP, statistical results from the
focal plane show that the average intensities of the two orthogonal
directions are non-identical ($A_1\approx0.554$ and
$A_2\approx0.446$), because the number of the element of MRPCP in
our real calculations is finite. Theoretically, the shape of the
probability density distribution depends on the ratio of the two
average intensities $A_1$ and $A_2$. For comparison, we directly
calculate the probability density of MRPCP+RPP from
Eq.~(\ref{Nonequal}), and draw it in Fig.~\ref{Fig6} (blue line). It
indicates that the probability of zero-intensity is zero and the
maximum probability moves toward the higher intensity. This
theoretical results are well in accordance with the statistical ones
(indicating by green squares in Fig.~\ref{Fig6}). The speckle
contrast is also reduced and determined by Eq.~(\ref{Contrast}),
here, $C=0.711$. The reduction of the speckle contrast is about
28.9$\%$ which is approaching to the ideal value 29.3$\%$.

Actually, incoherent laser irradiation are required by many
applications fields, such as in laser display, laser fusion, laser
heating processing, and some spectroscopy experiments. Our scheme
holds great potential in these applications, especially enables to
produce incoherent irradiation at very small dimensions (e.g., in
the order of wavelength) due to the subwavelength structure features
of metamaterial, which is not feasible for previous methods. An
example is fabricating such structure directly at the tip of an
optical fiber to form a compact device for beam
decoherence~\cite{Feng2011}.

\section{Conclusions}
We have proposed a metamaterial-based random polarization control
plate for eliminating the coherence of laser and producing
incoherent laser irradiation, since metamaterial can offer degrees
of freedom to tailor their transmission and polarization properties
by locally changing its structure geometry. A random polarization
control plate based on L-shaped plasmonic metamaterial has been
designed, which has locally varying structure parameters from which
the desired far-field polarization distribution can be obtained. The
calculating results show that it can effectively suppress the
speckle contrast and increasing irradiation uniformity, which is
potential to satisfy the increasing requirements for incoherent
laser irradiation and especially enables beam decoherence at very
small dimensions.

\begin{acknowledgements}
The author X. Ling sincerely thanks the anonymous referees for their
valuable suggestions. This research was partially supported by the
National Natural Science Foundation of China (Grant Nos. 61025024
and 11074068).

\end{acknowledgements}

\end{document}